\documentstyle[aps,pre,preprint,epsfig,eqsecnum]{revtex}
\input epsf

\begin{document} 

\title{A Novel Method for Sampling Alpha-Helical Protein Backbones}
\author{Boris Fain and Michael Levitt}
\address{Department of Structural Biology, 
Stanford University School of Medicine, 
\\  Stanford, CA 94305}
\maketitle
\begin{abstract}

We present a novel technique of sampling the configurations of helical proteins. 
Assuming knowledge of native secondary structure, we employ assembly rules 
gathered from a database of existing structures to enumerate the geometrically 
possible three-dimensional arrangements of the constituent helices. We produce a library 
of possible folds for twenty-five helical protein cores. In each case, our method finds significant numbers of 
conformations close to the native structure. In addition we assign coordinates to 
all atoms for four of the twenty-five proteins and show that this has a small effect 
on the number of near-native conformations.  In the context of database driven 
exhaustive enumeration our method performs extremely well, yielding significant percentages of 
structures (between 0.02\% and 82\%) within \mbox{6\AA} of the native structure. The method's 
speed and efficiency make it a valuable tool for predicting protein structure. 

\end{abstract}

{\em keywords}: proteins; helices; decoys; core. 

\section{Introduction}
\subsection{Outline of current prediction methodology} 
Prediction of protein structure from sequence is one of the most enticing 
goals of scientific inquiry today. Currently the most reliable method of determining a 
protein's shape is to search for a close homologue in the database of solved protein structures. 
Although the number of solved structures increases daily, it is estimated 
(Brenner {\em et al.}, 1997) that 
in the near future at least 40\% 
of proteins of interest bear no 
discernible sequence resemblance to a known macromolecule. Therefore {\em ab initio} prediction 
of structure from sequence remains an important challenge. 
 
When sequence homology cannot be used to construct 
a 3-D model, the current {\em modus operandi} for predicting 
structure is composed of three separate (yet interdependent) steps: sampling, searching, and 
ranking. 
First one picks a formalism to index and sample the possible structures, 
then searches through the conformation space and constructs the shapes, and finally uses some ranking criterion to pick out a structure as close to native as possible. Let us provide a few examples 
of each step. 
Some of the possible choices in the initial phase, picking a representation, are 
all-atom, reduced 'pseudo' amino-acid interaction centers
(Levitt \& Warshel, 1975; Levitt, 1976), lattice 
(Hinds \& Levitt, 1992; Covell, 1994; Lau \& Dill, 1989; Skolnick \& Kolinski, 1990), 
and many others.  Once 
the representation is chosen, one then picks the degrees of freedom to use. 
Researchers may vary Cartesian coordinates of all or 
a subset of the atoms, dihedral angles, relative distances, etc 
(Eyurich {\em et al.}, 1999). 
The decisions of which, how many, discretized or continuous, and so on,  are motivated by the amount of detail one wants to capture, as 
well as by computational complexity of the forthcoming search. 

Having picked a representation, one then selects an appropriate search technique, 
or a mixture of search techniques to march through the enormous space of 
possible 3-dimensional structures. As we have mentioned above, the three 
complementary steps of prediction are highly interdependent. The  most important 
factors in picking a search method are computational complexity and the 
natures of both the energy function and the representation. Lattice models, for instance, lend 
themselves easily to exhaustive enumeration. All atom Cartesian coordinate 
representations usually employ a physical potential, and search for a minimum 
with either an adaptive minimization routine, or a copy of Nature's 
algorithm, molecular dynamics. 

The remaining tool employed in protein structure prediction is a potential 
function constructed to pick out native or near-native conformations from a 
vast number of alternatives. If the search method employed is a guided one 
(for example, minimization) then the potential function also provides 
a landscape that will allow the procedure to converge to the correct answer. 
Just like sampling and search techniques, potential functions are many and varied: 
physical 
potentials with/without water representation, knowledge based potentials of 
various detail, hydrophobic contact potentials, and many
others 
(Sippl, 1990; Samudrala, 1997; Park \& Levitt, 1996). Frequently more than one 
potential is used to evaluate candidate structures. 

\section{Description of the Sampling Method}
\subsection{Graph-Theoretical representation of secondary structure}
We represent alpha-helical proteins as connected graphs. Each helix is
represented by a vertex and an edge is drawn between two vertices if the 
corresponding helices are in contact. To form a protein core one builds 
a connected subgraph, adding one helix at a time, until the protein 
is assembled. Figure \ref {myo_graph_fig} shows the graph-theoretical 
representation and the construction of Myoglobin.

We are, in essence, building an off-lattice model for secondary structure 
segments, but our degrees of freedom are not in the frequently chosen 
(Park \& Levitt, 1996; Eyrich {\em et al.}, 1999) 
$\left(\phi,\psi\right)$ loop residues (or their subsets), but in the 
relative geometric position of the helices themselves. The disadvantage of 
this approach is that a significant fraction of visited structures violate 
loop constraints - simply put, the ends of helices are too far apart to be 
joined together by an intervening loop. There are, however, several advantages. 
First, the structures we generate have a much higher tendency to be compact. 
Second, we are able to exploit the correlation between the sequence patterns
of helix-helix contact to significantly enrich our sampling with native-like 
structures. (A simple example of this is two helices with small residues in 
the contact area are more likely to be close to each other than those with 
large residues between them.) Third, we are able to sample possible packing 
much better than when using loop torsion angles, as is usually done in off-lattice 
models. 

\subsection{Exhaustive enumeration of helix-helix contacts}
\begin{figure}
\begin{center}
\leavevmode
\epsfxsize=6in \epsfbox{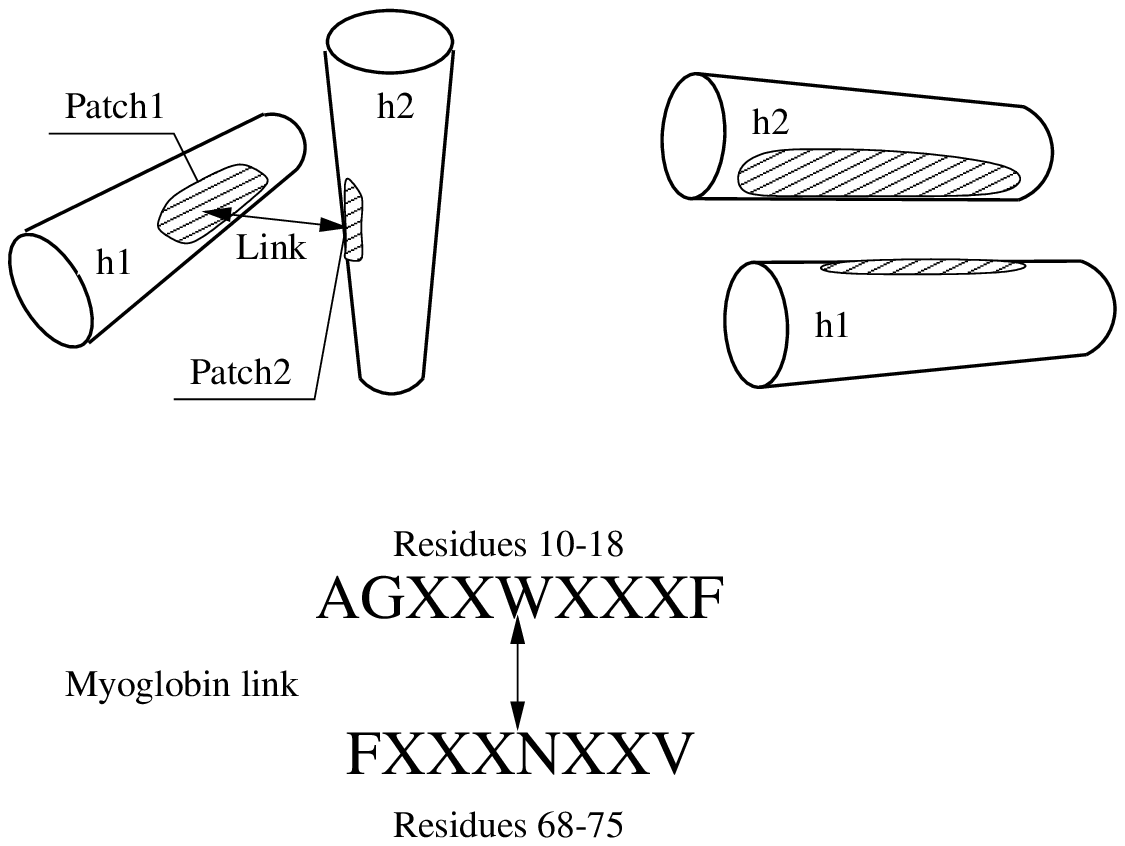}
\end{center}

\caption[]{\label{link_fig}
A diagram of two typical link configurations. The amino-acids in the 
shaded areas are involved in sequence matching against the target. The 
configuration on the left produces shorter patches than the configuration 
on the right. The figure on the bottom shows the sequence content of the
link. This is an actual link taken from Myoglobin {\em 1mba}. }
\end{figure}

In this work our principal aim is to achieve coarse-grained sampling of the 
helical protein core. 
 Our hypothesis is that 
the contact patches on each helix influence the way the helices pack. 
We choose to derive the preferences for helix-helix packing from 
the distribution of such orientations in known proteins.
If we are incorrect, 
then our packing strategy will position helices randomly and no harm will be done.
Therefore, the way we choose to travel through our space is by 
exhaustive enumeration of a discrete version of our representation. We can choose 
to sample relative helix-helix orientation in several ways. (We call a particular 
realization of an edge a 'link'. Figure \ref{link_fig} shows the various attributes of a link). 
It's useful, once again, to  
draw an analogy to dihedral sampling: there one can discretize a particular 
dihedral angle into a (not necessarily) uniform spectrum of values suggested 
by geometric considerations; e.g. sampling a dihedral angle in, say, 30 degree 
intervals. Alternatively, one can extract the local moves from a database 
of existing structures, thus sampling the space more efficiently. One could, for example, 
assign three possible values - helix, loop, sheet - to a dihedral 
angle.  Another possibility would be to bias the assignment 
with sequence matching.  
(Simons {\em et al.} 1997) 
or secondary structure prediction. 
In this work we choose exhaustive enumeration of possible orientations derived 
from known structures. This is motivated by the observed influence sequence has
on the packing of proteins 
(Reddy \& Blundell, 1993; Popov, 1980; Matheson \& Sheraga, 1978; Richmond
\& Richards, 1978). 

\subsection {Definition of helix to helix contact ('link') and the contacts' database}
We extract the possible relative orientations of two helices from a 
database of touching 
helix-helix pairs obtained from a subset of the SCOP 
(Murzin {\em et al.},1995)
database. The 1305 folds in our library have a sequence identity to each other 
no greater than 35\%. 
We then parse our database to get touching helix-helix pairs.  
We define two helices $X$ and $Y$ to be in contact if a) the shortest distance between 
any two CB's, $CB_{x}$ and $CB_{y}$, located respectively on helix $X$ and helix $Y$, 
is less than 7.3 \mbox{\AA}; and b) if each 
helix has at least 3 CB atoms within $d = d_{min} + 2.5$ \mbox{\AA}. \footnote{The three CB on each 
helix are necessary to define a relative rotation from one helix to another.}
The specific values of $d_{min} = 7.3$\mbox{\AA} and 2.5\mbox{\AA} worked best in our tests. Lowering the ${\em d}_{min}$ parameter
to below 5\mbox{\AA} caused some of the proteins in our database to be represented 
by disconnected graphs, thus making it impossible to ever reproduce them with our 
technique. Conversely, making the ${\em d}_min$ cutoff much larger than 7.3$\AA$ decreases 
the influence the contact sequence has on the relative orientation of the 
helices. 
In addition to the minimal three CB's on each helix, all the CB's on both helices that are within 2.5 \mbox{\AA} of $d_{min} = 7.3$ 
are defined to be in a contact 'patch'. 
Our definition picks out the residues in the contact region by 	
assigning a 'patch' of contact residues between helices. Residues on the 
far side of each helix, which have very little influence on the relative 
orientation, are ignored. Figure \ref{link_fig} shows the patches and the structure 
of each link in the database.

\subsection {The Enumeration Procedure}

\begin{figure}[hbt]
\begin{center}
\leavevmode
\epsfxsize=6in \epsfbox{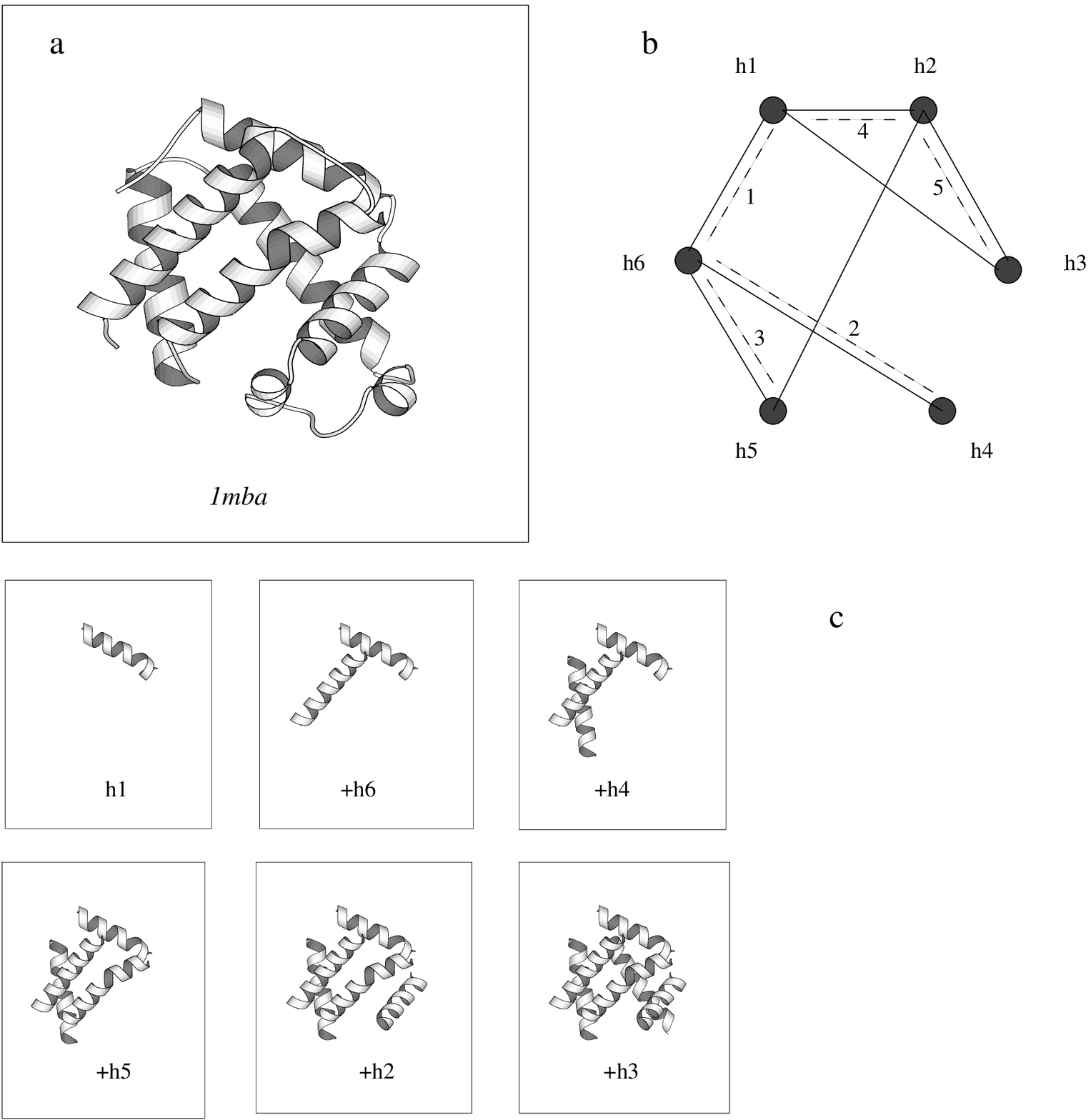}
\end{center}
\caption[]{\label{myo_graph_fig} This figure shows the graph representation 
of the protein Myoglobin, as well as its construction by our method. 
a) 1mba. This well known globin consists of 6 major and two minor helices. 
We sample the 6 major ones. 
b) The vertices of the graph are helices 
1 through 6 of 1mba. (helix1: residues 4-19; h2: 21-35; h3: 59-77; h4: 81-98; 
h5: 102-119; h6: 126-144.) The vertices represent the 
6 major helices; the solid lines are drawn whenever two helices are, 
according to our definition, in contact. The minimal spanning subgraph, 
outlined in dashed lines, represents one possible way to reassemble the helices. 
The dashed lines are numbered in the order the protein is assembled in 
one of our conformations.
c) The order of assembly of the helical core. Helices 
are assembled in the order 1, 6, 4, 5, 2, 3 using the relative orientation 
from the dashed lines of b). 
Figures were made with the aid of MOLSCRIPT (Kraulis, 1991).

In the process of construction other 'links' may form, however 
they are not neccesary to reproduce the helical core.}
\end{figure}

\begin{figure}
\begin{center}
\leavevmode
\epsfxsize=3in \epsfbox{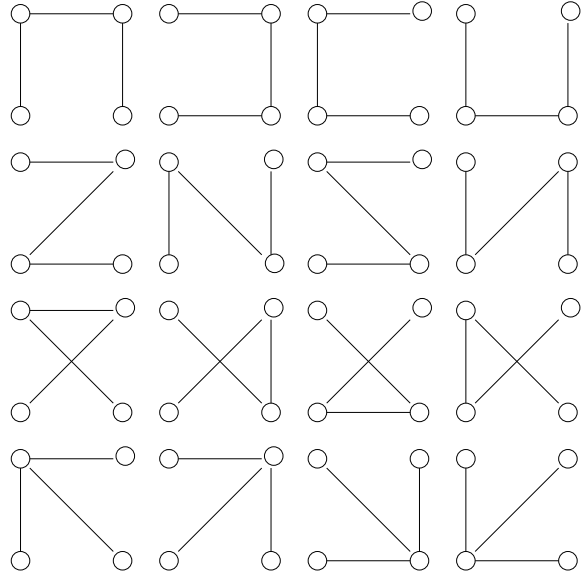}
\end{center}

\caption[]{\label{4h_graph_fig}
This figure shows the 16 topologically distinct ways to construct a 4-helix 
protein. In general there are $N^{\left(N-2\right)}$ spanning trees for a \mbox{N} vertex graph. }
\end{figure}

Figure \ref {myo_graph_fig} illustrates the enumeration procedure. 
The build-up of a helical protein proceeds as follows: given a protein sequence and native secondary 
structure assignment, we construct idealized helices on the chain. (In a 
predictive scenario the native secondary structure assignment will be replaced by a prediction, 
or possibly several alternative predictions.) We then go to the library of links 
and perform a sequence alignment between the patches on each link and all possible 
pairs of helices on our target sequence. The residues between key 'patch' residues serve as spacers to fix the position 
of the influential patch residues. 
If the sequence match score is high then this 
particular link will be used to bring the pair of helices together. The actual 
threshold for matching depends on the number of structures we want to sample. 
The sequence match is scored using the Blosum62 
(Henikoff \& Henikoff, 1992) matrix. 
The comparison threshold typically varies from 
0.1 to 0.6 and is adjusted to give anywhere from 10 to 1000 possible links for 
a given pair of helices in the target protein. The number chosen depends on how 
many final structures we wish to generate in the available amount of computer time. 
The relative orientations of the helices are then loaded into memory and the 
buildup of structures begins. \\
We use each topologically distinct pathway to build the target - (refer to figure 
\ref {myo_graph_fig} b) ) - each specific pathway corresponds to a minimal subgraph spanning 
the protein graph. (Dashed and solid lines, respectively, in figure \ref {myo_graph_fig}).
For a structure consisting of 4 helices this results in $4^{(4-2)} = 16$ possible topologies (see 
Figure \ref{4h_graph_fig}). Note that the graph 
corresponding to each final structure might (indeed should) possess other edges 
not used in construction; however it's sufficient to follow a minimally connected 
subgraph to construct it.

For each topology we construct each possible combination of links (gathered 
from the sequence-matching procedure above) that can realize a given edge. 
To improve performance we used branch and cut filters for loop and clash 
constraints. The clash filter eliminates a conformation if more than 3 residues 
on one helix are closer than 2 $\AA$ to residues on another helix. The loop 
filter eliminates conformations for which the distance for the loops necessary 
to connect the helices is longer than the maximum available loop length. The reason 
for the branch-cut approach is simple: if in a given six-helix enumeration helix one clashes 
with helix three there is no need to cycle through and build helices four, five, 
and six. Finally, each geometrically viable structure is tested for 
compactness. All the tests and filters are extremely fast because 
whenever possible we use the coarse segment representation of the structure 
and thus escape having to visit each amino acid's coordinates. We have 
also incorporated other filtering information, most notably disulfide bond 
locations, into the build-up procedure. 
In its current incarnation the method is able to generate roughly $10^3$ 
conformations per second on a 400MHz Pentium workstation. The ultimate speed of the procedure 
will, in the future, be limited by scoring function evaluations.

\section{Results}

To test the performance of our technique we have used the coordinates of 
helical cores for 25 proteins. The molecular sizes range from 31 to 
172 amino acids, and the number of helical residues to which 
coordinates are assigned ranged from 23 to 130.

The position of disulfide bonds can sometimes be easily obtained through 
chemical methods, and one can then rely on knowing their location prior 
to prediction of structure.  To this end, for 4 of the 25 proteins in 
our set that have disulfide links, we have also evaluated the 
performance of our method with and without {\em a priori} knowledge  
of these bonds. 

The fact that we assign residues only to the helical backbone leaves open 
the possibility that our results will be degraded by subsequent assignment 
of coordinates to all remaining atoms. We investigated this effect by 
building all atomic coordinates for 5 of our proteins using the program SegMod 
(Levitt, 1992). Prior to reconstructing full coordinates we pruned the helical cores with a slightly 
modified Sippl-like function 
(Sippl, 1990), leaving 500 to 4500 structures for 
each protein. (We left more decoys for larger proteins). 

\subsection{Sampling of helical cores}

We used the method to generate conformations for 25 helical proteins, ranging in size 
from 31 to 172 residues and containing anywhere from two to six helices. 
The program STRIDE 
(Frishman \& Argos, 1995) was used to determine the \
secondary structure of the native protein. 
To check how much the quality of our sampling will be degraded when all 
the coordinates are reconstructed from the helical core, we have included 
several proteins with significant loop content. 
The assignments of helices to the structure were made identical to the native 
structure. 

The results are summarized in table \ref{act_ss_results}.
Overall the results are very promising. For nearly every 	
protein a sizeable proportion of the sampled structures is within 
3\mbox{\AA} of the native. For 9 proteins the best structure produced is 
closer than 1\mbox{\AA} CA RMSD to the actual structure, and is virtually identical 
to the protein itself. Our procedure clearly samples well enough for a suitable potential 	
to make a successful prediction. 	

Our method is not only very fast, it is 
also very efficient, in the sense of being able to generate a large 
percentage of native-like conformations.  
We can access the efficiency by estimating 
how many random protein-like compact structures are needed to obtain the 
RMSD of our best structure, and compare that number to the number of conformations
we have visited. We use the estimate from 
(Reva {\em et al.}, 1998) 

\begin{equation}
\frac{1}{\left(\sigma\sqrt{2\pi}\right)}\int_{-\infty}^{R} \;
\exp\left[\frac{-\left(x - <R>\right)^{2}}{2\sigma^{2}} \right]dx
\label{reva_formula_eqn}
\end{equation}
Where, following Reva {\em et. al.}, 1998, 
we set $\sigma = 2.0$ and $<R> = 3.333N^{1/3}$, 
where $N$ is the number of residues in the protein core that we have assigned 
coordinates to. The proximity of our best structures to native exposes 
the weakness in estimating probability of low RMSDs, namely that 
the Gaussian distribution cannot be used to effectively describe sets of 
conformations which are very close to native. Having said that, we still 
feel that equation \ref{reva_formula_eqn} gives a good estimate of efficiency. 
The last column in table \ref{act_ss_results} shows the ratio of the number of 
structures we sample to the number of random structures given by formula 
\ref{reva_formula_eqn}. 

The results from this column in table \ref{act_ss_results} show that our method is efficient - 
the sampling is strongly biased towards the native conformation. 
The worst performance is shown in protein 1res, where we would have done 
slightly better if we chose relative helical conformations at random. However 
for most cases we are besting the random sampling by factors of $10^4$ to 
$10^6$, with efficiency actually improving for large proteins - which is 
where it matters most. 

\subsection{Features of the sampled conformations} 
\begin{figure}[hbt]
\begin{center}
\leavevmode
\epsfxsize=6in \epsfbox{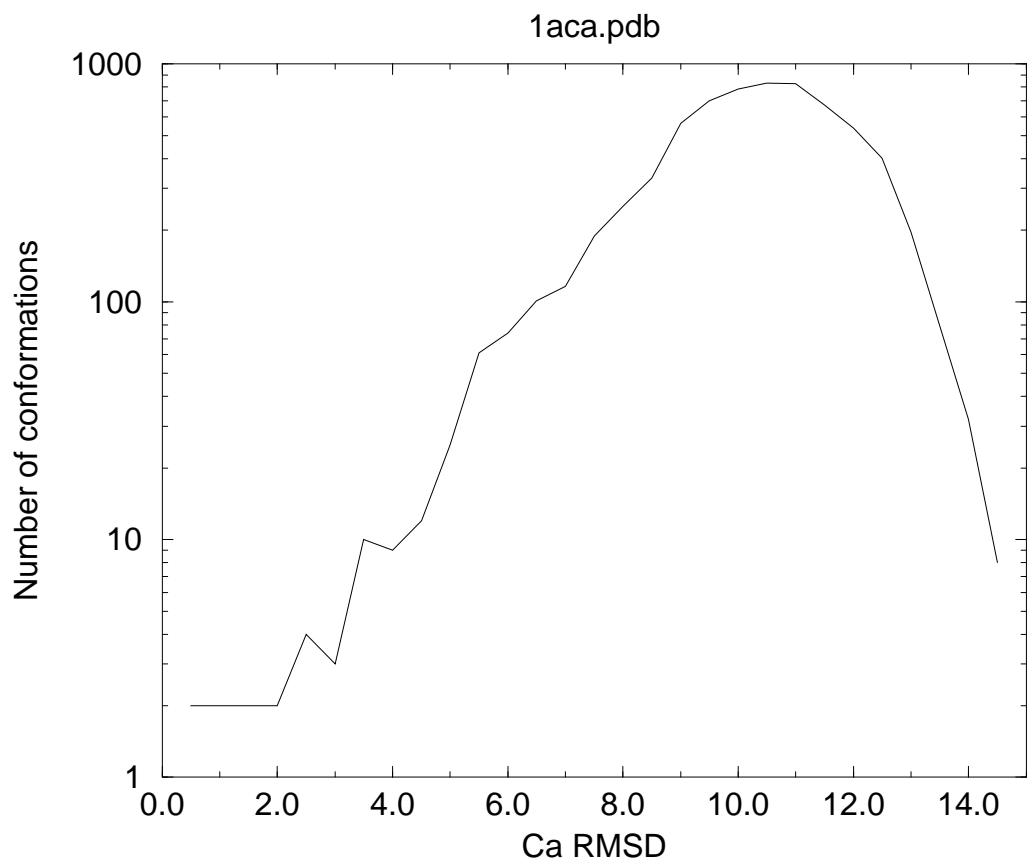}
\end{center}
\caption[]{\label{aca_hist_fig}
RMSD from the native for sampled conformations for 1aca.pdb. 
The distribution appears gaussian with an enriched tail of native-like 
structures}
\end{figure}

Figure \ref{aca_hist_fig} shows a typical distribution of the sampled 
conformations. One particularly interesting feature is the non-Gaussian 
tail extending towards the native structure. This is most likely the 
result of using sequence matching to select the possible orientations 
of the helix-helix pairs. The use of sequence information enriches 
the set with native-like structures and makes the curve decidedly non-Gaussian. 

\subsection{Using disulfide bond information}
To test how well the sampling is aided by supplementary information, we 
took proteins with disulfide bonds and subjected the sampling for those 
structures to an additional filter. We did not screen 1cc5 
because although it does have disulfide bonds, they occur in the loop 
regions and do not help filter the helical cores. Our filters rejected 
conformations in which the CYS CB-CB distance was greater than 8\mbox{\AA}. 
This is not a very strict filter; it simply weeds out grossly incorrect 
topologies. 
The results both with and without (The results 
without disulfide bond information are the same as in table 
\ref{act_ss_results}) disulfide bond information are 
displayed in table \ref{disulfide_table}. The number of structures 
was cut by approximately a factor of 8. While the filtering did not significantly improve 
the average RMSD of the ensemble, the proportion of 'good' structures 
was raised approximately fivefold. 

\subsection{Completing the structure}

The structures described in the previous two sections possess only backbone 
coordinates for residues in the helical core. It is quite possible that the 
successes of the sampling 
can be washed out when coordinates are assigned to the 
remaining residues and side-chain atoms. To check how the 
quality of the sampling changes during completion we ran our sets of
conformations through the program SegMod (Levitt, 1992). 
We did not construct full coordinates for all twenty five proteins because we simply wanted to 
show that reconstruction does not significantly degrade the quality of the 
sampling procedure. To make the test credible, we selected proteins of varying size, and 
with long unassigned loops. The results are shown in table \ref{complete_table}. \\ 
Once again, the resulting ensembles of decoys contain a sizeable proportion of near-native 
decoys. Even though we did not produce any decoys closer than 6$\AA$ for 2fha, we feel that 
the sampling of such a large (according to {\em ab initio} 
prediction standards) protein - 172 residues - was successful.  The RMSD of the best decoy, 7.3$\AA$ 
has log-odds of 23234 (computed from last column of table \ref{complete_table}.
The total ensemble for 2fha has 0.25
results are good for decoy ensembles. Whether they are good enough for prediction remains to be seen. 

\section{Discussion}

\subsection{Desirable features of search methods} 

In this paper we presented a novel sampling technique for helical proteins. 
To provide a context in which one can judge the merits of our procedure, 
we shall review the features that make a sampling method effective. 

Our assertion is that for a search method 
to be useful in {\em ab-initio} structure prediction, 
the proportion of sampled conformations that are native-like should be statistically significant. 
Admittedly, the definitions of both ``statistically significant'' and ``native-like''
are imprecise - but clearly, the more and the closer, the better. 

The need to get close to the native fold is motivated by several considerations.  
First, we wish to furnish good, i.e. close to native, predictions, Since we cannot 
select that which we do not have, having at least one native-like conformation is an 
obvious necessity. 
Second, an energy functions can only be effective in a slice of conformational space 
close to the actual protein . Structures can be 'wrong' in a vast number of ways, 
many of which are low scoring alternatives to the native fold in a very large space. Only a small 
subspace close to the actual structure can be expected to exhibit the properties 
of an energy funnel; i.e. a region where the energy decreases uniformly as one 
approaches the native, and where a majority of conformations are more 
favorable than all the others in the available space. We can demand that an energy 
function correctly discriminates a {\mbox 1 \AA} fold from all {\mbox 10 \AA} folds, but 
if our best candidate model is {\mbox 8 \AA} cRMS from native, we cannot expect an 
energy function to pick this over one that is {\mbox 10 \AA} away. 
The need to have more than one 'good' structure; indeed to enrich the library as much 
as possible with near-native choices, is brought about by the imperfections of 
the available selection methods.  A perfect energy function should be able to
discriminate a
native structure from {\em all} other alternatives - this,
in fact, is what
nature is able to do with striking consistency.

The current generation of potentials (Samudrala \& Moult, 1997; Sippl,
1990, Simons {\em et al.}, 1999)
often misidentifies incorrect folds, giving them energies 
lower than that of the native fold. By 
increasing the number of 'correct' answers we increase the statistical likelihood of 
making a successful prediction. In addition, some current techniques use consensus 
information - selecting a subset and building a consensus model from it 
(Huang {\em et al.}, 1999). 
These methods clearly need a sizeable native-like population to work.
To give today's energy functions all the help they can certainly use, it's advantageous 
to have as many native-like folds as possible in one's decoy set. 

To make the above discussion more concrete, we should quantify the definition of 
being 'near-native. How close is close enough? One possible answer comes from 
comparing one's technique to 
random sampling. Reva {\em et. al.} (1998) 
suggest a 'native-like' 
target value of about \mbox{6 \AA} cRMS deviation relative to the actual structure. 
Other authors 
(Park \& Levitt, 1996; Simons {\em et al.}, 1997). 
also propose a similar distance of \mbox{4-6 \AA}. 
\footnote{This number depends on the size of the protein.} 

\subsection{Comparison with other search methods}
Our sampling method has both merits and drawbacks when compared with other
search techniques. Many classical approaches of assembling alpha-helices
into a core have been proposed 
(Soloviev \& Kolchanov, 1981; Cohen {\em et al.}, 1979, Cohen \& Sternberg,
1980). 
One alternate approach to sampling
is to assemble helices using distance geometry by either embedding
distance space in three 
dimensions
(Aszodi {\em et al.}, 1995; Havel, 1991), or by minimizing against restraints 
(Cohen \& Sternberg, 1980; Kuntz{\em et al.}, 1982; Richmond \& Richards,
1978; Huang {\em et al.}, 1999; Chelvanayagam {\em et al.}, 1998; Lund {\em
et al.}, 1996; Mumenthaler \& Braun, 1995; Ortiz {\em et al.}, 1998;
Smith-Brown {\em et al.}, 1998). 
In a recent work 
(Huang {\em et al.}, 1999) this method produced decoy libraries with the 
log-odds of producing the lowest structure in the -4 to -6 range
\footnote{Log-odds are $1/N_{best}$, where $N_{best}$ is the number of random structures 
needed to find at least one structure with RMSD equal to our best structure. $N_{best}$ 
is the first number in the last column of table \ref{act_ss_results}.}.
Our results, which range from log-odds of -4 to -9, are  somewhat better. However we must 
emphasize that the comparison is unfair, since we used actual structure, and Huang et. al. 
used predicted (albeit well predicted) secondary structure. We are currently examining 
how well our method will perform with predicted secondary structure.  Preliminary indications 
are that both distance geometry and our method both get sufficiently close to the native fold, with 
our method having a slight edge due to its speed. 

Our method begins to pull away from distance geometry methods in the scaling of CPU time needed 
when one moves to larger and larger proteins.
Current distance geometry methods cannot be extended to proteins of length over 
100 residues 
(Huang {\em et al.}, 1999). In contrast, because our sampling technique scales 
with the number of secondary structure segments and not the number of residues, it can 
easily handle chains of 100 - 200 residues, comparable to single domains of larger proteins.
The library for the largest protein in our work - 2fha, 172 residues long - took approximately 15 hours of CPU time on 
a 400 MHz Pentium II machine. The sampling procedure produces approximately 500 structures per second 
for two-helical proteins, 100 per second for proteins containing three
helices, 10 per second with four, and 
roughly one structure/second for proteins with five and six helices.

Another sampling technique is to fix the helical segments and then to vary the dihedral 
angles of the loop regions. This method has been used to produce fold
libraries 
(Park \& Levitt, 1996),  
and, combined with a branch and bound algorithm, in folding studies 
(Eyrich, {\em et al.}, 1999).
This method carries an advantage over ours because every structure generated automatically 
satisfies loop constraints, where as many of our potential structures do not. Many of our 
structures have to be rejected because the physical distance between ends of helices violates 
chain connectivity. Furthermore, we do not 
have loops at the end of construction; therefore we are currently working on an algorithm 
to sample loop regions of our protein cores that would be a good (i.e. fast) match to our 
helical core sampling.

Our method out-paces the loop dihedral angle sampling in two categories: scaling of 
computational demands with time, and efficient sampling of structures with plausible contacts. 
Let us illustrate the first difference with a four-helical protein. Assuming an average 
loop length of seven residues varying loop conformation demands 7(residues)*2($\phi, \psi$)*3 = 42 
degrees of freedom. Our procedure needs to position three rigid bodies (the position of the initial 
helix is arbitrary), requiring 6(rigid body)*3 = 18 degrees of freedom. 
The second difference stems from the fact each helix in our construction method is guaranteed 
to have at least one, and possibly two plausible contacts with other helices. When one 
varies dihedral angles, most of the non-clashing structures have large voids between helices. 

Yet another method for generating possible reduced-model conformations for protein structure 
are lattice and off-lattice models, such as 
Hinds \& Levitt, 1992, 1994; Covell, 1994. 
These methods are better ours in their generality because they require no knowledge of either actual or 
predicted secondary structure. On the other hand it's difficult to see secondary structure 
at the resolution of these lattice models. In addition, often secondary 
structure is fitted onto the conformations at the end of construction. 
In computational performance, for small proteins these approaches compare very favorably 
with ours - a simplified representation which assigns 
a lattice point to every second residue can exhaustively sample shapes of proteins of up to 
100 residues 
(Hinds \& Levitt, 1992). However, because of an exponential increase in the number of shapes 
of a self-avoiding walk on a tetrahedral lattice, it's difficult at this point to see a 
generalization of the lattice methods that would apply to larger molecules without a significant
sacrifice in resolution. 

One final method we want to mention is that of assembling structures from fragments of existing 
folds 
used by Simons {\em et al.}, 1997, 1999. We find this method very appealing. It uses the information 
in the sequence to bias the assembly much the same way we use the 'patch' information to bias 
helix-helix orientation. It is reported to produce best structures as good as the lowest RMSDs 
in our set. In addition, this method - as do the lattice models and the loop angle search - 
works on beta and mixed alpha/beta proteins: something we cannot yet do. 
Because we have not had the opportunity to test this method ourselves, we cannot 
comment on how its efficiency compares with ours. Our feeling is that sequence information 
in 'contact patches' has more correlation on the global geometry of the
structure than the local sequence/structure combination employed in these
methods. Also our method is much faster than any of the other currently 
available sampling methods.

\subsection{Advantages and disadvantages}

In the beginning of his section we introduced two desirable qualities of a sampling 
method: speed, and the ability to produce structures close to the native fold. 
Our technique fills both of these requirements. 

The speed of the method depends on 
the size of the protein and, more specifically, on the number of helices 
that we are trying to arrange. 
Each entry in table \ref{act_ss_results} takes from a fraction of a second 
to a few hours to produce. A naive count would estimate a factorial 
growth in the number of graphs, and an additional geometric growth 
in the number of structures sampled in each graph; however the actual 
increase is much less. For most structures of 4 helices or more, cutting 
the branches that violate self-avoidance results in a significant 
reduction of the number of conformations we need to sample. In our tests 
each additional helix increased the time of the runs by approximately 
a factor of 30. The set of conformations for our largest structure, 1fha, took approximately 
4 hours to generate. 

In addition to being fast, our method scales well - the use of
branch-cutting helps because larger proteins are very constrained by 
self-intersection. Because of this our technique 
is able to sample molecules comparable in size to small domains, which 
are the largest single-chain structures one wishes to predict. 

We chose to re-arrange helices using a 'patch' contact database derived from 
existing structures. This approach significantly enriches the sampled ensemble 
with native-like structures. In addition, the graph-theoretical enumeration 
of relative orientations ensures that we sample all plausible regions of conformation space. 

In its current incarnation our method also possesses some drawbacks. The main 
disadvantage of our technique is the requirement of a specified secondary structure. 
Ideally a search technique should also sample alternate secondary 
structure assignments. Currently we can only do this by specifying different 
assignments at the beginning of the procedure. 
Another significant drawback is the absence of a complimentary loop-building 
method. We need a method with the speed of a loop-library lookup methods, 
yet able to get a 3 \mbox{\AA} or better approximation 
of the native loop (so that our near-native cores remain near-native). 
A final challenge to our approach is the lack of an obvious generalization to 
beta and alpha/beta proteins. The definitions of a sub-segment and the contact 
patch have to be significantly revised to adapt to beta sheets.

\section{Future directions} 

The deficiencies of our technique, outlined in the previous paragraph, point 
the way to future developments. We are currently working on a fast 
loop building procedure for short (3-10 residues) loops. 
Our next project is to enhance sampling of alpha proteins to include 
variations of the boundaries of the helices. 

A slightly more distant goal is the 
extension of the presented technique to construct $\beta$-sheet and mixed 
${\alpha/\beta}$ proteins. In 
addition we are working on a fast preliminary discrimination function which 
would be used prior to the reconstruction of all atoms for each conformation.

\section{Acknowledgements}
Many thanks to the present and former members of the Levitt group for 
help and discussions. We wish to thank Patrice Koehl for providing 
excellent routines for superimposing structures. 
Thanks to Prof. Joseph Rudnick for hints on graph theory. B.F. wishes 
to thank the A.P. Sloan Foundation, and the U.S. Department of Energy for financial 
support. This work was supported in part by grant DE-FG03-95ER62135 to M.L.
from the U.S.Department of Energy.

\begin{table}[hbt]
\caption{\label{act_ss_results}Summary of helical core sampling.}
\begin{tabular}{lcclllc}

Name&Helices&Assigned /&CA RMSD range&\% below 3\mbox{\AA}&\% below 6\mbox{\AA}&Random /\\
& &Total residues&best - worst : ave& & &Generated\\
\tableline
T0065&2&23/31&0.483 - 7.503 : 4.988&3.1\%&81.4\%&291299/2604\\
1fc2 C&2&23/43&0.759 - 7.937 : 4.813&10.3\%&77.4\%&153634/2144\\
1res&3&24/43&2.265 - 9.646 : 6.546&0.055\%&29.3\%&8392/9028\\
1erp&3&26/38&0.986 - 9.052 : 6.191&0.26\%&41.8\%&226495/11188\\
1mbh&3&30/52&0.954 - 10.412 : 7.123&0.3\%&23.3\%&773292/11724\\
1uxd&3&30/59&0.880 - 10.828 : 6.773&0.58\%&28.2\%&927522/16281\\
3hdd&3&40/56&0.240 - 12.022 : 8.001&4.0\%&11.8\%&8.3E7/13380\\
1trl&3&42/62&0.617 - 12.920 : 8.002&0.24\%&12.4\%&4.0E7/24151\\
T0073&2&43/48&0.662 - 8.348 : 4.840&11.9\%&78.9\%&5.5E7/522\\
1cc5&4&43/83&3.484 - 13.237 : 9.241&-&0.84\%&47664/46725\\
1r69&5&44/63&3.127 - 13.143 : 9.140&-&1.1\%&128078/56520\\
1lfb&3&44/77&1.584 - 13.668 : 8.973&0.09\%&5.4\%&5.68E6/41505\\
2ezh&4&48/65&3.514 - 15.363 : 9.670&-&1.4\%&5E6/61981\\
1c5a&4&49/65&3.774 - 13.731 : 9.064&-&1.7\%&78745/98842\\
1hsn&4&50/79&3.321 - 17.380 : 11.211&-&0.28\%&266620/32734\\
1ropa&2&51/56&2.432 - 13.751 : 6.765&1.3\%&40.2\%&2.9E6/620\\
1pou&4&52/71&4.546 - 14.318 : 10.187&-&0.95\%&25296/29283\\
1nre&3&57/81&2.240 - 13.020 : 8.215&0.21\%&11.3\%&1.6E7/21716\\
1ail&3&60/70&3.028 - 15.481 : 9.582&-&2.5\%&3.6E6/15980\\
1nkl&4&60/78&3.879 - 14.122 : 9.931&-&1.15\%&439759/15486\\
1aca&4&60/86&0.752 - 15.059 : 10.442&0.13\%&1.9\%&2.5E9/6827\\
1flx&4&67/79&3.306 - 14.786 : 10.869&-&0.635\%&6.5E6/37141\\
1aj3&3&88/98&2.594 - 16.309 : 10.125&0.05\%&3.1\%&1.6E9/8177\\
1lis&5&91/131&4.345 - 18.696 : 12.430&-&0.07\%&2.0E7/32805\\
2fha&5&130/172&4.855 - 21.671 : 15.912&-&0.023\%&1.0E10/4279\\
\end{tabular}
\end{table}
Table \ref{act_ss_results} summarizes the sampling of helical cores for 25 proteins. 
Column 3 shows the number of residues that are assigned coordinates as well 
as the total length of the protein. Column 4 lists the range of CA RMSD of 
assigned coordinates from native coordinates. Columns 5 and 6 show the 
percentage of total structures that are closer than, respectively, 3$\AA$ 
and 6$\AA$ to the native structure. The last column displays the efficiency 
of our decoy generation method. It shows the ratio of the number of structures one needs to generate randomly in order to produce the best 
CA RMSD  (computed using Eqn. \ref{reva_formula_eqn}) to the number of 
conformations in our ensemble. 
\\

\begin{table}[hbt]
\caption{\label{disulfide_table}Disulfide bond information included}
\begin{tabular}{lcccc}
Protein&ave without $\rightarrow$ with&\%below 3A&\%below 6A&num/numact\\
\tableline
1c5a&9.064 $\rightarrow$ 8.047&-&1.7\% $\rightarrow$ 7.4\%&98842 $\rightarrow$ 6325\\
1erp&6.191 $\rightarrow$ 6.017&0.26\% $\rightarrow$ 1.3\%&41.8\% $\rightarrow$ 46.9\%&11188 $\rightarrow$ 761\\
1nkl&9.931 $\rightarrow$ 9.254&-&1.15\% $\rightarrow$ 5.8\%&15486 $\rightarrow$ 889\\
\end{tabular}
\end{table}
Table \ref{disulfide_table} displays the results of pruning the decoy ensembles with 
disulfide bond information. The columns show shifts in the average Ca RMSD, proportions 
with RMSD below 3$\AA$ and 6$\AA$, and the reduction in the number of structures in 
each set. 
\begin{table}[hbt]
\caption{\label{complete_table}RMSD's for full structures}
\begin{tabular}{lclllc}
Name&Size, Helices&CA RMSD range&\% below 4\mbox{\AA}&\% below 6\mbox{\AA}&Random/\\
& &best - worst : ave& & &Generated\\
\tableline
T0073&48, 2&1.435 - 8.320 : 4.840&24.5\%&73.7\%&2.14E7/522\\
1ropa&56, 2&2.446 - 15.041 : 7.074&6.6\%&34.0\%&7.79E6/712\\
1ail&70, 3&3.200 - 15.624 : 9.730&0.26&3.50\%&1.45E7/1917\\
1flx&79, 4&3.282 - 14.886 : 11.329&0.06&0.9\%&5.56E7/3342\\
1aj3&98, 3&2.67 - 16.02 : 10.396&1.0\%&4.0\%&9.17E9/1880\\
2fha&172, 5&6.644 - 26.323 : 18.668&-&0.25 \%$\leq 10\AA$&7.27E8/4279\\
\end{tabular}
\end{table}
Table \ref{complete_table} summarizes the results for decoy sets with 
all atoms' coordinates assigned. The columns are similar to table 
\ref{act_ss_results}. The number of decoys in each ensemble is smaller 
than in table \ref{act_ss_results} because of pruning with a statistical 
pairwise potential. 
2fha had no conformations below 6$\AA$, however 
0.25 \% of structures were closer than 10$\AA$ to the native.

\end{document}